\documentclass[5p,twocolumn]{elsarticle}

\usepackage{amsmath,url}
\usepackage[switch]{lineno}
\usepackage{natbib}

\usepackage{multirow}

\journal{Astroparticle Physics}

\begin{document}

\begin{frontmatter}

\title{Calibration of the Cherenkov Telescope Array using Cosmic Ray Electrons}

\author[rvt]{R.D. Parsons}
\author[rvt]{J.A. Hinton}
\author[rvt]{H. Schoorlemmer}
\address[rvt]{Max-Planck-Institut f\"ur Kernphysik, P.O. Box 103980, D 69029, Heidelberg, Germany}

\date{\today}

\begin{abstract}
Cosmic ray electrons represent a background for gamma-ray observations
with Cherenkov telescopes, initiating air-showers which are 
difficult to distinguish from photon-initiated showers. This similarity,
however, and the presence of cosmic ray electrons in every field
observed, makes them potentially very useful for calibration
purposes. Here we study the precision with which the relative energy scale
and collection area/efficiency for photons can be established using
electrons for a major next generation instrument such as CTA.
We find that variations in collection efficiency on hour timescales
can be corrected to better than 1\%. Furthermore, the break in the
electron spectrum at $\sim$0.9 TeV can be used to calibrate the energy scale
at the 3\% level on the same timescale. For observations on the order
 of hours, statistical errors become negligible below a few TeV and allow for
an energy scale cross-check with instruments such as CALET and AMS. Cosmic ray electrons therefore provide a powerful calibration tool, either as an alternative to intensive
atmospheric monitoring and modelling efforts, or for independent
verification of such procedures.
\end{abstract}

\begin{keyword}
Electron-positron spectrum, Imaging Atmospheric Cherenkov Telescopes,
calibration, CTA, gamma rays 
\end{keyword}

\end{frontmatter}

\section{Introduction}
\label{sec:intro}

Electrons (and positrons) represent $<$1\% of the cosmic ray flux at
100 GeV energy. However, after the hadron-rejection cuts typically
applied to date taken by
Cherenkov telescope arrays, they represent a dominant
background over a wide energy range, with improving hadron rejection
compensating for the steeper electron spectrum ($\sim E^{-3}$ versus
$\sim~E^{-2.7}$) up to the break in the electron spectrum
at 900~GeV~\cite{HESS_electrons_PRL}. The electron background is uniform on
the sky at the $<$5\% level below 100 GeV~\cite{Fermi_electrons}, %and expectations
while at higher energies the anisotropy is unknown (although
anisotropy is expected to increase with energy).
Electrons are therefore present in every field observed by
Cherenkov telescope arrays, with close to isotropic flux, and
separable from protons and nuclei using modern
background-rejection methods~\cite{HESS_electrons_PRL,MAGIC_RF,stefan_tmva, Mattieu_model,Yvonne}.
Once the electron spectrum is known, the rate and spectrum
measured in a given observation can be used to correct for atmospheric
and instrumental deviations from the ideal case, or to check that
atmospheric and instrumental corrections have been successfully applied. The advantages over
cosmic ray protons and nuclei for this purpose (see for example~\citep[][]{Jamie_and_Stephan}) are
the close similarity of gamma and electron initiated air showers in terms of
morphology and atmospheric depth at which the maximum number of particles is reached, albeit with a half radiation length
shift, and the presence of a distinct feature in the CR electron
spectrum: the 0.9~TeV break. This feature raises the prospect of
independently establishing collection area and energy scale changes,
something which is impossible using single power-law spectra. The
spectral break position
and level of high energy anisotropy in
electrons will be established independently by future ground-based
Cherenkov telescope arrays and by space-based
instruments such as CALET~\cite{CALET} and perhaps AMS~\cite{ SpecAMS2014},
providing a means for cross-calibration of the instrument based on a independent energy scale.

Measuring the cosmic ray electron spectrum with an array of Imaging Atmospheric Cherenkov
Telescopes (IACTs) is, however, a significant challenge. The
H.E.S.S. collaboration was the first to demonstrate that this is at all
possible, by applying hard selection cuts (four telescope multiplicity and a
random forest approach)~\citep{HESS_electrons_PRL}. Subsequently, these measurements
were extended to lower energies for H.E.S.S.~\citep{HESS_katrin} and now confirmed by 
MAGIC~\citep{SpecMAGIC2011} and VERITAS~\citep{SpecVERITAS2015}. For current-generation instruments these
measurements require long exposures: typically many hundreds of hours.
Spectral measurements for gamma-ray sources make use of background
estimates established using regions in the field of view thought to be
empty of gamma-ray emission. This approach is clearly not possible for
electrons, which are close to isotropic. Instead a model of the
background in terms of some separation parameter (for example the
output of a neural network classifier) must be established. This
requires a detailed understanding of the development of hadronic
cascades in the Earth's atmosphere. Significant
differences exist (at the $\sim10\%$ level) between hadronic interaction models (or
Monte-Carlo event/interaction generators) due to underlying physical
uncertainties, particularly in the production of pions with a large
forward momentum in the energy range of interest \cite{ConstraintsLHC,SystematicsAir}.
Dedicated instruments at the LHC, such as LHCf and TOTEM, as well as the general purpose ATLAS and CMS detectors, have now
significantly reduced the uncertainties in this energy range and models such
as EPOS LHC and QGSJETII are currently being refined to reflect these
developments~\cite{QGSJETII,EPOS_LHC}. The systematic uncertainties on electron spectrum
extraction will therefore be much smaller in the near future than
those presented in the existing IACT publications. 

The next generation facility CTA (the Cherenkov Telescope Array~\cite{CTA_CONCEPT})
will employ over 100 telescopes at two sites (CTA-North and CTA-South), dramatically improving
on the performance of current generation IACTs.
The wider field of view of CTA telescopes ($\sim$8$^{\circ}$ diameter), lower
energy threshold ($\sim$20 GeV), and very large collection area of the instrument (typically an order of magnitude larger
than current instruments for gamma ray analyses, and even more for
electron analyses due to the hard cuts often used, at all energies)~\cite{CTA-MC-Prod2} combine to produce an electron
rate after quality selection cuts that is two or more orders of magnitude larger
than that measured by current arrays~\cite{Dan_thesis} at $\sim$0.9 TeV. Furthermore, the background
rejection power of CTA will be superior to that of current generation
instruments, allowing the extraction of the cosmic ray electron
spectrum over a wide energy range in a short time, with modest
systematic uncertainties~\citep{Dan_thesis}. 

CTA will employ LIDAR-based atmospheric monitoring systems to measure variation in light propagation through the atmosphere (\citep[]{CTA_CALIB}, and references therein). Whilst these measurements will be used to ensure realistic atmospheric treatment in the Monte Carlo simulation of the detector response, it is highly desirable to have a procedure for continuous confirmation that
such measurement procedures have been successful, and as an independent means of
deriving correction factors.
In addition, instrumental effects may
change the efficiency with which gamma-like showers trigger the array
and pass selection cuts, and/or lead to systematic under or over
estimation of photon energy. Again, CTA will make use of multiple
methods to characterise such effects, but the approach of deriving the
cosmic ray electron spectrum in a routine way for all observations
without a significant diffuse gamma-ray component
promises a convenient end-to-end method to establish correct performance
or to derive correction factors. Due to the lack of bright diffuse
gamma-ray emission in the relevant energy range and the small angular
size of point-like sources compared to the instrument field-of-view,
the electron spectrum can be extracted from almost all potential CTA
extragalactic observations without the addition of an gamma-ray
electron separation, simply by the removal of significant point
sources from a given observation set (typically 1 source per field in
the current generation of telescopes).

Here, we propose a method for a CTA electron spectrum measurement and
assess the timescales on which the flux normalisation and break energy
can be found.
We go on to discuss the systematic uncertainties
associated with this approach and its merits for the array-level calibration of CTA.

\section{Approach}
\label{sec:approach}

To test the feasibility of using the electron spectrum as a means of
high level calibration, electron spectral measurements were simulated
using the  CTA-South
``Production-2'' Monte Carlo dataset~\cite{CTA-MC-Prod2}.  Array layout ``2Q''
was used, which contains 4 large sized telescopes (23\,m diameter), 24 medium
sized telescopes (12\,m) and 72 small sized telescopes
(4\,m). Direction and energy reconstruction were performed using the
CTA baseline analysis\footnote{Consisting of Hillas parameterisation of images and a weighted combination of the intersection of image axes for direction reconstruction, and energy estimation using look-up tables~\cite{BernlohrCTA2013}.}, under the assumption that the events are 
diffuse electrons. To ensure the quality of the images that
are used in the reconstruction, we apply cuts on the number of
pixels and number of photo-electrons (p.e.), these selection criteria
were optimised for the nominal night-sky background rate
(extrapolated from measurements at the H.E.S.S. site)  
and are summarised in table 1 for each telescope type. To
improve the quality of the reconstructed air shower parameters we
require that the reconstructed shower direction lies within 4$^{\circ}$ of the
telescope pointing direction and that a minimum of four telescopes participated in
the reconstruction.

 \begin{figure*}[t]
\begin{center}
\includegraphics[width=0.49\textwidth]{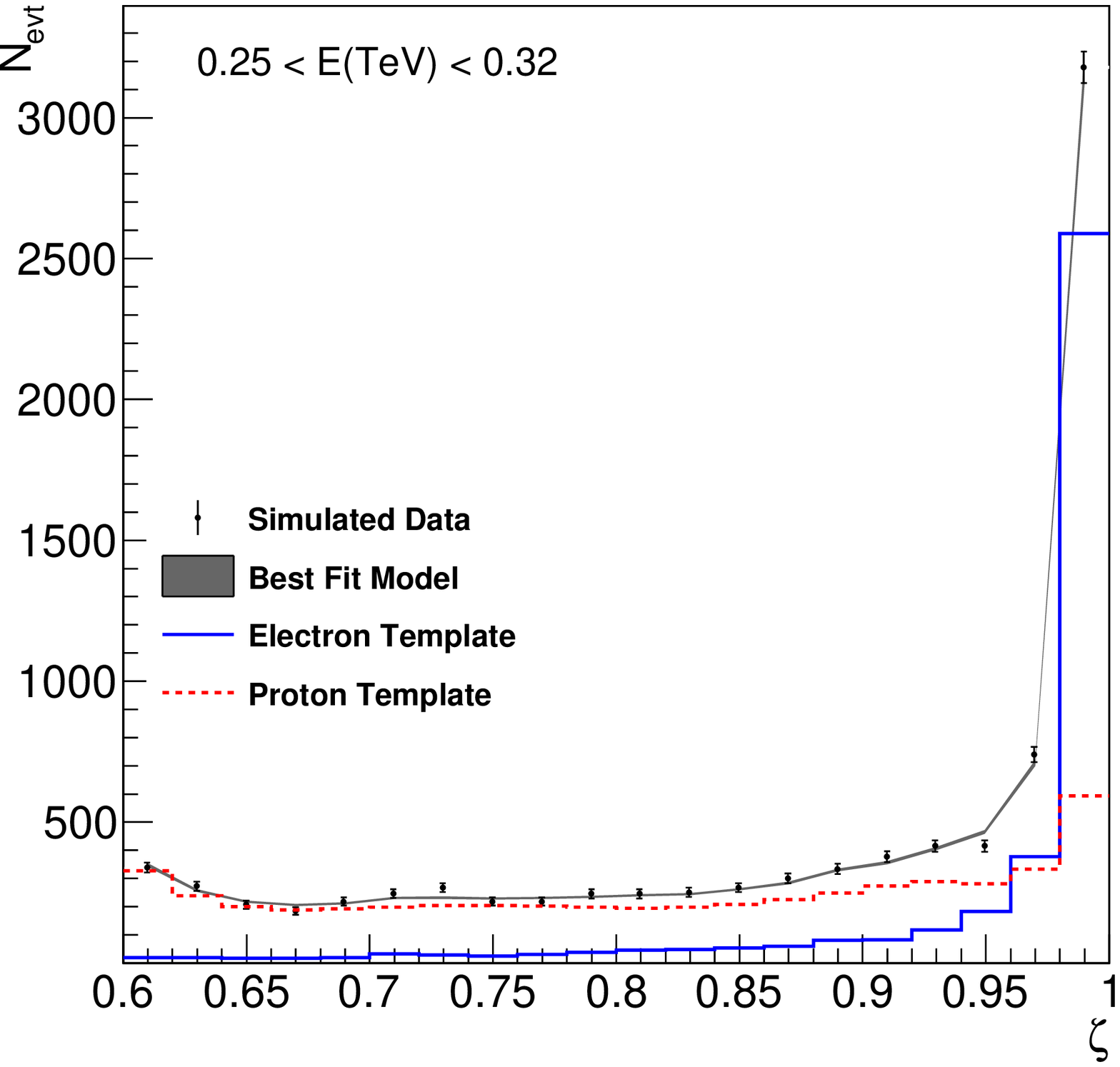}
\includegraphics[width=0.49\textwidth]{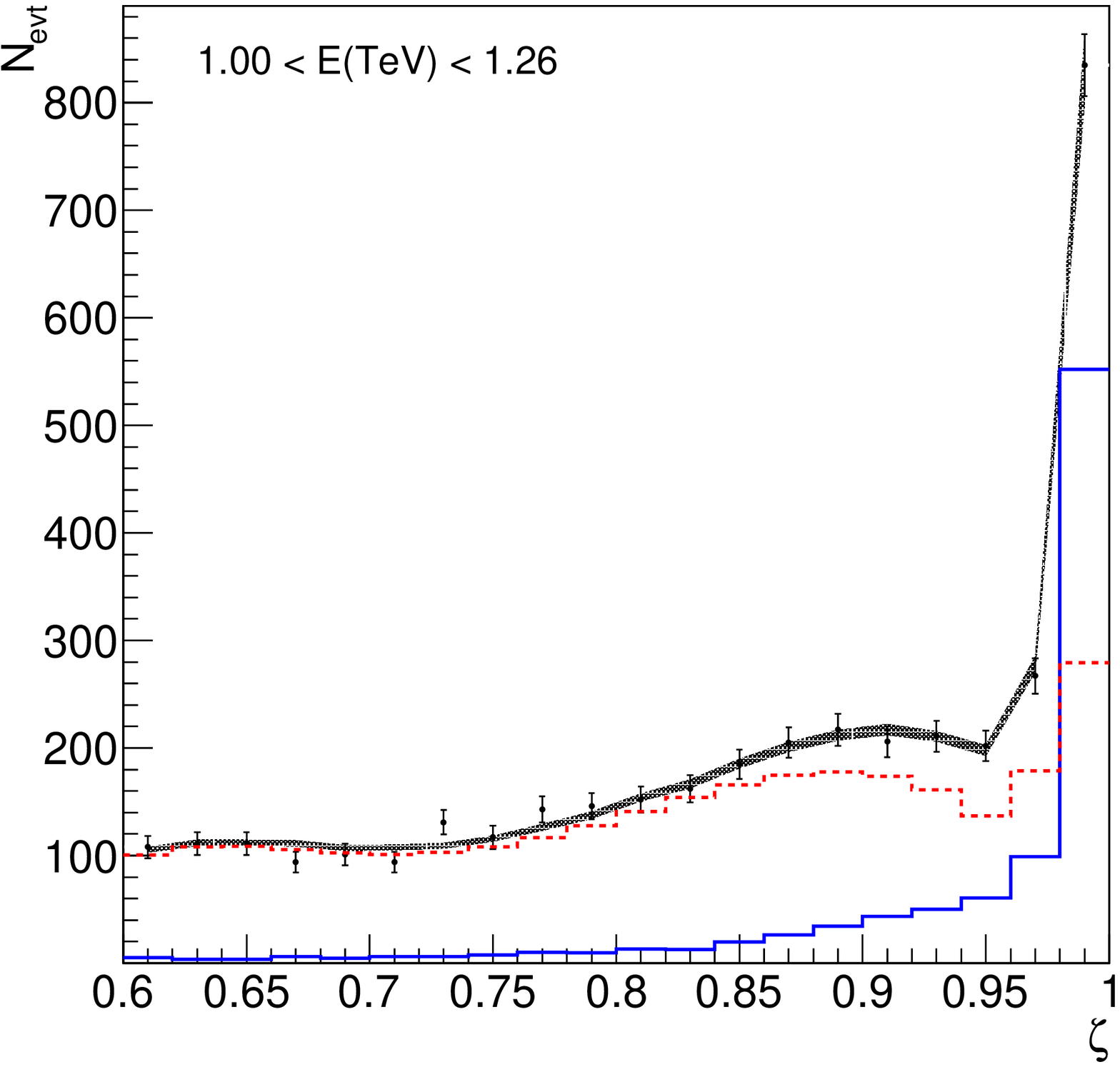}
\caption{Simulated distributions of the electron separation parameter
  $\zeta$ for an hour observation of a $\gamma$-ray source free
  region with CTA at an energy interval below (left) and above the
  (right) break $E_b$ in the spectrum (proton distribution smoothed by
  polynomial fitting). The 1-$\sigma$ error band of the best fitted model is also indicated.}
 
\label{fig-distribution}
\end{center}
\end{figure*}

An artificial neural network was created using the TMVA package
\cite{TMVA2007} to perform classification of electrons from
protons. The neural network was trained in five energy bins covering
the full energy range of the CTA instrument (0.02--100~TeV), using the
following discriminating variables:

\begin{itemize}
\item[-] Mean scaled event width/length (see e.g. \cite{HESSCrabPaper})
\item[-] Root mean square of scaled event width/length between telescopes
\item[-] Root mean square of event energy estimates between telescopes
\item[-] Reconstructed depth of shower maximum ($X_{\text{max}}$)
\item[-] Spread of $X_{\text{max}}$ estimates between telescopes
\item[-] Mean time gradient across an image \cite{TimeGradient}
\end{itemize}

\begin{table}[htp]
\begin{center}
	\label{tab-telCuts}
	\begin{tabular}{|ccc|}
	 	\hline	
		Type & Amplitude (p.e.) & N$_{\text{pix}}$ \\
		\hline
		Large (4) &  $>$92.7 & $\geq$5 \\ 
		Medium (24) &  $>$90.6 & $\geq$4 \\
		Small (72) &  $>$29.5 & $\geq$4 \\
		\hline
		\end{tabular} 
	\caption{Image cuts for the different type of telescopes.
        }
	\end{center}
\end{table}

Once trained, an independent sample of simulated data was passed through the network to 
produce the expected classifier ($\zeta$) distributions of electrons and protons.
Combining these distributions with the correct normalisations to 
provide the expected distribution of events when observing a gamma-ray free region of the sky requires assumptions on the spectra of protons and electrons,
for which we adopt the following functional form for protons (based on
data from \cite{proton_spec}):

\begin{equation}
  F_{p} = \phi_{0,\rm{p}} \left( \frac{E}{1\,\mathrm{TeV}} \right)^{\Gamma} %\mathrm{m}^{-2}  \mathrm{s}^{-1}  \mathrm{TeV}^{-1} \mathrm{sr}^{-1}
\end{equation}
with $\phi_{0,\rm{p}} = 9.6 \times 10^{-2} \, \mathrm{m}^{-2}  \mathrm{s}^{-1}  
\mathrm{TeV}^{-1} \mathrm{sr}^{-1}$ and 
$\Gamma = -2.7$. 
For electrons we have
\begin{equation}
F_{e} = \phi_{0.\rm{e}} \left( \frac{E}{1\,\mathrm{TeV}} \right)^{\Gamma_{1}}
\left[ 1+\left(\frac{E}{E_{b}} \right)^{\frac{1}{\alpha}}\right]^{(\Gamma_{2} - \Gamma_{1})\alpha} 
\end{equation}
 with $\phi_{0,\rm{e}} = 1.5 \times 10^{-4} \,\mathrm{m}^{-2}  \mathrm{s}^{-1}
\mathrm{TeV}^{-1} \mathrm{sr}^{-1}$, 
$\Gamma_1 = -3.0$,
$\Gamma_2 = -4.1$,
$E_b = 0.9 \, \mathrm{TeV}$, and $\alpha = 0.2$ (the H.E.S.S. measurement
gives a limit of $\alpha < 0.3$), 
consistent with measurements using Fermi-LAT \cite{Fermi_electrons}, AMS \cite{SpecAMS2014} and
H.E.S.S.\cite{HESS_katrin} respectively.
The contribution of heavier cosmic-ray nuclei can be safely ignored, due
to their lower expected fluxes and the extremely powerful background
rejection for such events. This ``data'' distribution can then be scaled
and Poisson fluctuations added to represent any length of
observation time.  Once this simulated observation expectation has
been created, we use a component fitting technique similar to that
used in \cite{HESS_electrons_PRL}, using the aforementioned particle
classifier distributions to estimate the contribution of each particle
type to a given energy bin (see Figure~\ref{fig-distribution}). The number of electrons in a given bin
($N_{\rm{elec}, i}$) can then be estimated by integrating the 
fitted electron component. Figure~\ref{fig-distribution} gives
examples of simulated and fitted $\zeta$ distributions, we selected
one example below and one example above the energy at which the spectral break occurs.

\begin{figure}[t]
\begin{center}
\includegraphics[width=0.49\textwidth]{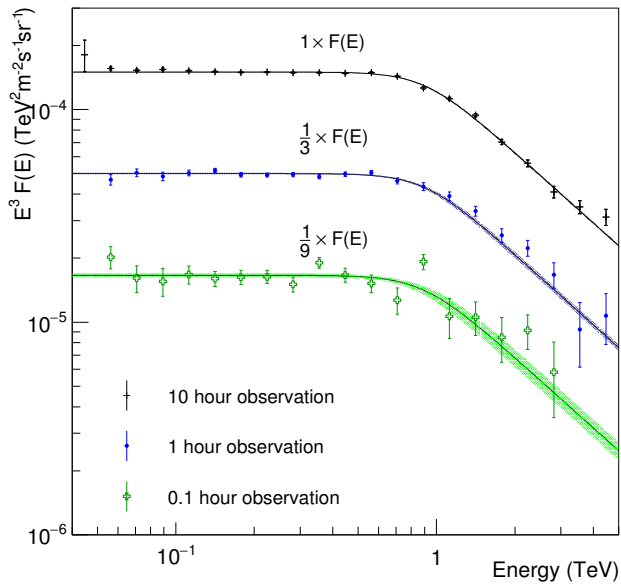}
\caption{Reconstructed electron spectra from simulations with CTA
  observations of different durations (scaled by different factors
  for readability). The markers illustrate the result from Monte Carlo
  simulation of a single observation run, while the lines and the
  shaded area indicate the average and the 68\% confidence interval of
  the fitted spectrum from many Monte-Carlo realisations.}
\label{fig-spec}
\end{center}
\end{figure}

Spectral fitting was performed using a forward folding technique
\cite{ffold}, utilising the full energy migration matrix and effective
area to produce the expected number of counts for a given spectrum. A
minimisation was then performed using the MINUIT
package\footnote{\url{http://lcgapp.cern.ch/project/cls/work-packages/mathlibs/minuit/index.html}}
to find the values of the spectral constants where the expected
distribution best matches $N_{\rm{elec}}(E)$.
Figure~\ref{fig-spec} gives illustrative reconstructed electron spectra on different timescales.

\section{Calibration Accuracy}

To test the accuracy of the spectral reconstruction 1000 realisations
of the $\zeta$ distribution in each energy bin were created by adding
random Poisson fluctuations to the expected event distributions. A
spectral fit was then performed on each realisation, with the spectral
index before and after the break fixed, but with break energy $E_{b}$
and flux normalisation as free parameters. The standard deviation of the
distribution of each parameter is taken as the uncertainty of that
parameter for a given observation time.

\begin{figure}[t]
\begin{center}
\includegraphics[width=0.49\textwidth]{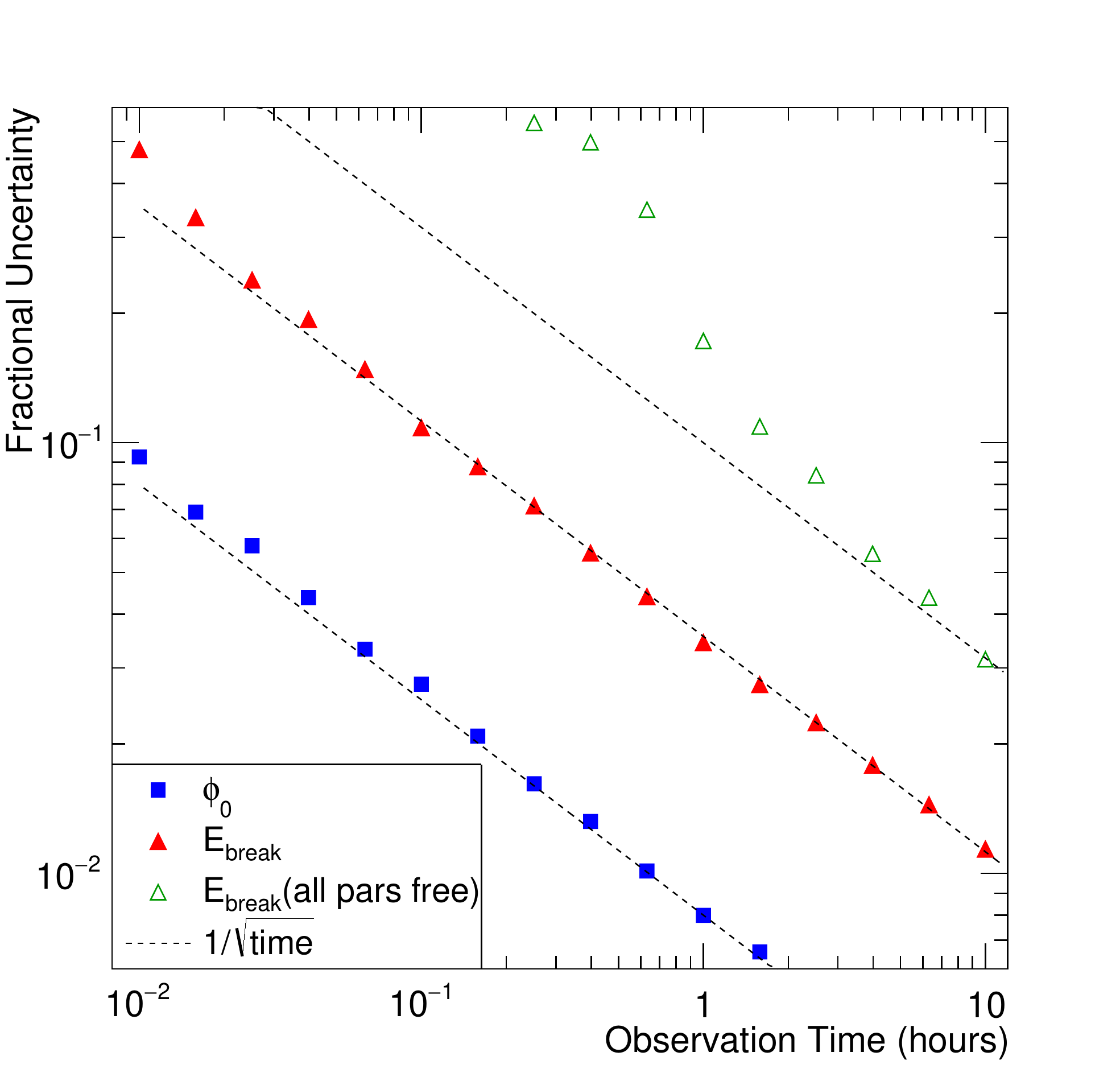}
\caption{
  Fractional statistical uncertainty on the reconstruction of spectral normalisation 
and energy break point as a function of observation
time. For the break energy both the case of fixed spectral index
before and after the break and the case with all parameters free are
shown. In the background limited regime a $1/\sqrt{\mathrm{time}}$
dependence is expected, illustrated by the dashed lines.
  \label{fig-accuracy}
  }
\end{center}
\end{figure}

Figure \ref{fig-accuracy} shows the evolution of the fractional
uncertainty in both the normalisation and break point of the spectrum
as a function of observation duration. One can see that the
uncertainty on the normalisation drops rapidly with time, reaching the
10\% level after only $\approx$1 minute. This rapid improvement in
accuracy is achievable as it is possible to calculate the
normalisation from only the low energy data, allowing high statistics
observations to be made in a relatively short time period. 

The accuracy of the measurement of the break point improves more
slowly, taking around 15 minutes to reach a 10\% accuracy. This
difference is due to the requirement of data points reconstructed
above 0.9~TeV in order to resolve the break. Figure \ref{fig-accuracy}
also shows the break point determination is all spectral parameters
are left free (used for example for the measurement of the electron reference
spectrum) the time taken to determine the cut-off to the 10\% level is
almost a factor of 10 larger, demonstrating the power of having a well
known spectral shape. Further study of the accuracy when using only
individual telescope subsystems, shows that the accuracy of this
calibration in the most part derives from the MST subsystem, reaching
10\% accuracy in almost the same time as the full system. This is most
likely due to the large effective area in the region of the energy
break of this subsystem. Whereas the LST subsystem with its much
smaller effective area takes around five times longer. While the
SST subsystem can take over ten times as long to calibrate to the 10\%
level, due to the threshold of this system being close to the energy
break point.
\label{sec:accuracy}

 It should be noted that the
current measurement of the electron spectrum has relatively large
systematic and statistical uncertainties at high energies, which leads
to fairly large uncertainties on the required timescales
for determination of the break energy. For example, if
the break energy is 20\% higher than assumed here, the required
observation time to reach a given accuracy would increase by a factor $\sim$3.

\section{Discussion}
\label{sec:discussion}

The results of section~\ref{sec:accuracy} indicate that rapid electron
spectral reconstruction should be possible with a next generation IACT
array, such as the Southern and Northern CTA arrays. However, it is important to consider the impact of systematics
on the electron spectral measurement and consequently the
use of cosmic ray electrons for calibration purposes. 

The model of calibration using this technique is first to construct a
reference electron dataset based on the best atmospheric quality
observations. This reference dataset can be used to generate a
spectral model and if significant anisotropy exists in the reference set,
or is established by CALET on this timescale, a directional model as
well. The model can then be compared to the nightwise spectrum, after
correction for atmospheric effects. This comparison provides an
important check of the stability of the CTA
spectral results over the lifetime of the array and under all atmospheric
conditions.

The absolute reference spectrum is affected by several systematic
effects, but in general the relative calibration factors determined by
comparison to the reference are much less susceptible to
systematics. The primary systematic uncertainties are expected to be:
\begin{itemize}
\item Uncertainties on the classifier distributions for proton
  showers, due to underlying uncertainties in {\bf hadronic
    interactions}. Such uncertainties will be reduced over the coming
  decades with data from the LHC, but are unlikely to become
  negligible.
\item {\bf Atmospheric uncertainties} in a given data set are of
  course one of the main targets of the electron calibration, but
  uncertainties in the average atmospheric conditions during best
  possible observing conditions lead to systematics in the reference
  spectrum. CTA plans for a thorough characterisation of the atmosphere
  and hence such uncertainties should be greatly reduced with respect
  to current IACTs.
\item {\bf Detector model / simulation uncertainties} can be reduced
  over time with careful monitoring as planned for CTA, and an
  iterative approach with the many calibration tools available. Such
  uncertainties can also affect the behaviour of the background
  rejection classifier (for example in regions of high night sky
  background noise) and care must be taken to ensure the stability of
  such classifiers using Monte Carlo simulations under different
  potential observing conditions.
\item Any significant cosmic-ray electron { \bf anisotropy} will
  clearly lead initially to calibration uncertainties for individual
  pointings, but once established can be modelled and accounted
  for. The only problematic case is small scale anisotropy at a level
  of several percent, something which is not to our knowledge
  predicted by any model.
\end{itemize}

Many systematic checks are possible to help understand any deviations
of measured spectra from the reference, or the reference spectrum from
space-based measurements. The available data can for example be split
into zenith and azimuth bands, into different telescope sub-systems
(large, medium and small size telescopes)
and different seasons, due to the very high statistics obtained
overall.

The advantages of this approach over more traditional methods are the
end-to-end nature of the calibration, sensitive to instrumental and
atmospheric changes, the similarity of electrons to gamma rays and
hence minimisation of systematics associated with extrapolation from
the rather different hadronic showers, and the spectral feature,
allowing collection area and energy-scale effects to be readily
separated. Additionally, due to the very high cosmic ray electron statistics
available from CTA it is possible to compare the electron spectrum reconstructed
from arbitrary sub-sets of telescopes, allowing further systematic
uncertainties to be probed (e.g. differing atmospheric absorption
across the array footprint). Although not directly investigated here
it should also be possible to perform such a calibration at the
northern CTA site, due to the dominance in the calibration of the MSTs
which CTA-North should have in similar numbers. 
The combination of data from ground-level muons with
electrons is particularly powerful, with muons providing a means to
calibrate individual
telescopes~\cite{whipplemuons,CTA_CALIB}, and electrons the
system as a whole. Both muons and electrons will be collected during
routine IACT observations and hence the downtime of the observatory due to
special calibration runs is minimised.

\section{Conclusions}
\label{sec:conc}

Tests of electron spectral reconstruction demonstrate that a measurement of
the relative normalisation and energy scaling factor of the cosmic-ray
electron spectrum is possible at the 3\% level with under one hour of CTA observations.
Such short timescale measurements make it
possible to use the electron background seen in all observation runs
as a ``standard candle'', allowing the systematics of the gamma-ray spectral
reconstruction of CTA to be assessed over the lifetime of the
instrument. This technique can be used in concert with the several
absolute calibration techniques for the atmosphere, individual
telescopes, and for the full array, that have been proposed for CTA \cite{CTA_CALIB}.
Together with point spread function verification using point-like gamma-ray sources (for example blazar flares), 
this procedure can be used as 
a high level check of the ``health'' of the instrument over its
expected 30 year operational lifetime, ensuring absolute data
corrections remain consistent and spectral results are stable. This conclusion can of course be extended from CTA to any instrument with substantially improved sensitivity and field of view with respect to current IACTs.

\section*{Acknowledgments}

This work was done as part of a major simulations and calibration
effort for CTA, and hence draws on the work and expertise of many
members of the CTA consortium. This paper has gone through internal
review by the CTA Consortium.

\end{document}